\documentclass[prb,final,twocolumn,showpacs,superscriptaddress,preprintnumbers,amsmath,amssymb,floatfix]{revtex4}

\usepackage[]{graphicx}

\begin{document}

\title{{\textbf{The charge ordered state in half-doped Bi-based manganites studied by
$^{17}$O and $^{209}$Bi NMR}}}

\date{\today}
\author{A.Trokiner}
\affiliation{ Laboratoire de Physique du Solide,
E.S.P.C.I.,Paris,France }
\author{S.Verkhovskii}
\affiliation{ Laboratoire de Physique du Solide,
E.S.P.C.I.,Paris,France } \affiliation{Institute  of Metal
Physics, Ural Branch of Russian Academy of Sciences, Ekaterinburg
GSP-170, Russia}
\author{A.Yakubovskii}
\affiliation{ Laboratoire de Physique du Solide,
E.S.P.C.I.,Paris,France }  \affiliation{Russian Research Centre
"Kurchatov Institute", Moscow, Russia}\affiliation{Division of
Physics, Graduate School of Sciences, Hokkaido University,
Sapporo, Japan}
\author{K.Kumagai}
\affiliation{Division of Physics, Graduate School of Sciences,
Hokkaido University, Sapporo, Japan}
\author{S-W.Cheong}
\affiliation{Department of Physics and Astronomy, State University
of New Jersey, Rutgers, USA}
\author{D.Khomskii}
\affiliation{II Physikalisches Institut, Universitaet zu Koln,
Germany }
\author{Y.Furukawa}
\affiliation{Division of Physics, Graduate School of Sciences,
Hokkaido University, Sapporo, Japan}
\author{J.S.Ahn}
\affiliation{Department of Physics and Astronomy, State University
of New Jersey, Rutgers, USA}
\author{A.Pogudin}
\affiliation{Institute  of Metal Physics, Ural Branch of Russian
Academy of Sciences, Ekaterinburg GSP-170, Russia}
\author{V.Ogloblichev}
\affiliation{Institute  of Metal Physics, Ural Branch of Russian
Academy of Sciences, Ekaterinburg GSP-170, Russia}
\author{A.Gerashenko}
\affiliation{Institute  of Metal Physics, Ural Branch of Russian
Academy of Sciences, Ekaterinburg GSP-170, Russia}
\author{K.Mikhalev}
\affiliation{Institute  of Metal Physics, Ural Branch of Russian
Academy of Sciences, Ekaterinburg GSP-170, Russia}
\author{Yu.Piskunov}
\affiliation{Institute  of Metal Physics, Ural Branch of Russian
Academy of Sciences, Ekaterinburg GSP-170, Russia}

\date{\today}

\begin{abstract}

We present a $^{209}$Bi and $^{17}$O NMR study of the Mn electron
spin correlations developed in the charge ordered state of
Bi$_{0.5}$Sr$_{0.5}$MnO$_{3}$ and Bi$_{0.5}$Ca$_{0.5}$MnO$_{3}$.
The unusually large local magnetic field $^{209}H_{loc}$ indicates
the dominant $6s^{2}$ character of the lone electron pair of
Bi$^{3+}$-ions in both compounds, probably, responsible for the
high temperature of charge ordering $T_{CO}$. The observed
difference in $^{209}H_{loc}$ for Bi$_{0.5}$Sr$_{0.5}$MnO$_{3}$ to
Bi$_{0.5}$Ca$_{0.5}$MnO$_{3}$ is suggested to be due to a decrease
in the canting of the staggered magnetic moments of Mn$^{3+}$-ions
from adjacent ab layers. The modification of the $^{17}$O spectra
below $T_{CO}$ demonstrates that the NMR line due to the apical
oxygens is a unique local tool to study the development of the Mn
spin correlations. In the AF state the analysis of the $^{17}$O
spectrum of Pr$_{0.5}$Ca$_{0.5}$MnO$_{3}$ and
Bi$_{0.5}$Sr$_{0.5}$MnO$_{3}$ points towards two different types
of charge ordering in these systems: a site-centered for the first
manganite and a bond-centered one for the second material.

\end{abstract}


\maketitle

\smallskip

\section{INTRODUCTION}

Among  doped transition metal oxides, one of the most
interesting systems to study  charge ordering (CO), orbital
ordering (OO), and spin ordering phenomena is the half-doped
manganites R$_{0.5}$A$_{0.5}$MnO$_{3}$ (R - rare-earth ion or Bi;
A - Ca, Sr). Furthermore, manganites show the colossal
magnetoresistance effect which results from the competition
between the insulating charge-ordered state and metallic
ferromagnetic state. Nevertheless, the role of charge ordering
and/or orbital ordering in the magnetotransport properties of
manganites remains unclear.

At high temperature R$_{0.5}$A$_{0.5}$MnO$_{3}$ is in the
charge-disordered (CD) paramagnetic (PM) state. The disorder is
related to the thermally activated hoping of  $e_{g}$-holes. This
hoping causes ferromagnetic (FM) correlations between the electron
spins of the neighboring Mn-ions,
\cite{Kajimoto_PRB58,Cheong_PRL81,Yakubovskii_PRB67} the
spin-fluctuation parameters being controlled by the
double-exchange mechanism proposed by Zener. \cite{Zener_PR82}
Below $T_{CO}$ the localization of the mobile holes induces
transition to the CO phase. The spatial ordering of electrons over
given Mn-orbitals leads to an excess in the kinetic energy that is
compensated by the elastic energy due to the cooperative
Jahn-Teller distortions in the MnO$_{6}$-octahedra sublattice. The
orbital ordering is related to the Jahn-Teller distortions.

In the present work, OO and CO have been studied in Bi- based
manganites. The critical temperature of charge ordering is $T_{CO}
= 325$ K for Bi$_{0.5}$Ca$_{0.5}$MnO$_{3}$ \cite{Bokov_PSS20} and
475 K for Bi$_{0.5}$Sr$_{0.5}$MnO$_{3}$.\cite{Garcia_PRB63} One
open problem is why the critical temperature of charge ordering
$T_{CO}$ in Bi-based manganites is so much higher  than in their
rare-earth counterparts. It was established for the rare-earth
manganites RE$_{0.5}$(Ca,Sr)$_{0.5}$MnO$_{3}$ that the
$e_{g}$-band width is controlled by the (Mn-O-Mn)-bond bending
which in turn depends on $<r_{A}>$ - the average radius of
(RE,A)-cations. A smaller $<r_{A}>$ narrows $e_g$-band and favors
the hole localization thus increasing $T_{CO}$. Recently it was
shown, \cite{Garcia_PRB63} that such correlation between $T_{CO}$
and $<r_{A}>$ is not observed in Bi-based manganites
Bi$_{0.5}$(Ca$_{1-x}$,Sr$_{x}$)$_{0.5}$MnO$_{3}$. While $T_{CO}$
increases smoothly with $<r_{A}>$ for $x \leqslant 0.4$ these
manganites belong to one of the $Pbnm$ space subgroups in the
CO-phase. The Sr-rich manganites ($x \geqslant 0.6$) belong to
$Ibmm$ space subgroup and show the opposite tendency, i.e.
$T_{CO}$ decreases with $<r_{A}>$.
\cite{Beran_ChemMatt4,Frontera_PRB68,Hejtmanek_JAP93} Thus, the
high values of $T_{CO}$ for Bi$_{0.5}$Ca$_{0.5}$MnO$_{3}$ and
Bi$_{0.5}$Sr$_{0.5}$MnO$_{3}$ cannot be properly explained
considering only the average buckling of the (Mn-O-Mn)-bonds. It
was suggested \cite{Garcia_PRB63} that an increase of 6$s^{2}$
character in the stereoactivity of the Bi$^{3+}$ lone electron
pair could favor the growth of $T_{CO}$ for $x > 0.5$ but the
mechanism of the influence of lone pairs, as well as the detailed
orbital structure of these pairs in different Bi manganites, have
not been clarified.

\begin{figure}
\centerline{\includegraphics[width=0.95\hsize]{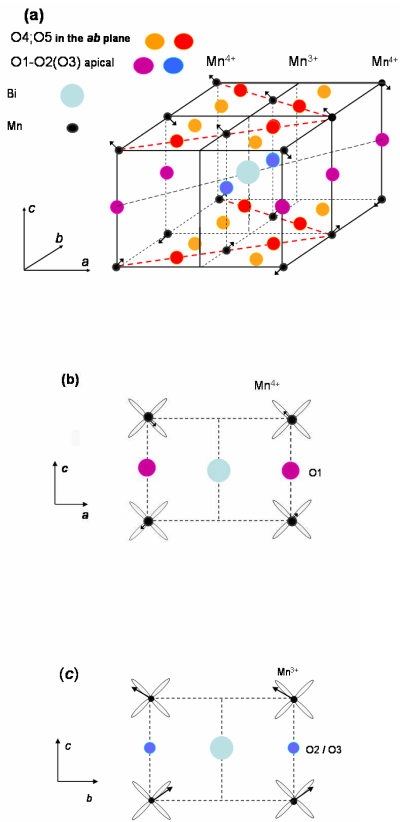}}
\caption{\label{fig_1} CE$'$- type magnetic structure of
Bi$_{0.5}$Sr$_{0.5}$MnO$_3$ \cite{Frontera_PRB64} represented in
the frame of a pure ionic model (site-centered order). The 3D
structure is represented in (a). The red dotted lines show the
ferromagnetically correlated zig-zag of Mn$^{3+}$ and Mn$^{4+}$
ions in the $ab$ planes. The environment of Bi cation in the $ac$
plane (b) and in the $bc$ plane (c) is represented with the
Mn-$t_g$ orbital orientation and the spin direction of the four
nearest Mn ions, Mn$^{+4}$ and Mn$^{3+}$, respectively. The (O1 -
O5) sites of oxygen are specified  similar to the notation
suggested in Ref.~\onlinecite{Jirak_PRB61}.}
\end{figure}

We have performed $^{209}$Bi zero-field NMR measurements on
Bi$_{0.5}$Ca$_{0.5}$MnO$_{3}$ and Bi$_{0.5}$Sr$_{0.5}$MnO$_{3}$ in
the AFM-ordered state. According to
Ref.~\onlinecite{Frontera_PRB64}, below the Neel temperature
($T_{N} \thickapprox 155$ K) Bi$_{0.5}$Sr$_{0.5}$MnO$_{3}$
exhibits a static spin order of the CE$'$-type
(see~Fig.~\ref{fig_1}). For Bi$_{0.5}$Ca$_{0.5}$MnO$_{3}$ in the
AFM state the magnetic structure ($T_{N} \thickapprox 140$ K) has
not been reported yet. For both compounds, we have observed a very
large value of $^{209}H_{loc}$, the local field at Bi. It depends
on the short-range spin order of the nearest Mn. For both samples
the analysis of $^{209}H_{loc}$ has been performed in the frame of
the CE$'$-type magnetic order, although it is not known for
Bi$_{0.5}$Ca$_{0.5}$MnO$_{3}$. Our results favors the $6s$
dominant character of the lone pair of Bi rather than the $6p$
one. To assertain magnetic structure of
Bi$_{0.5}$Ca$_{0.5}$MnO$_{3}$ the $^{17}$O NMR spectra were
measured in zero external magnetic field at low temperature.

The purpose of the $^{17}$O NMR study reported in this paper is to
trace the Mn-Mn electron spin correlations which develop when
cooling Bi$_{0.5}$Ca$_{0.5}$MnO$_{3}$ and
Bi$_{0.5}$Sr$_{0.5}$MnO$_{3}$ in the CD/CO PM phase. In the first
NMR investigation of the CO state of a half-doped manganite,
\cite{Yakubovskii_PRB67} it was shown that the $^{17}$O NMR probe
is a powerful tool to study the orbital ordering and the spin
ordering. Compared to other techniques, NMR is a local probe,
which enables us to obtain valuable information on the spin
correlation of Mn-ions through the orbital hybridization of Mn
with the probed oxygen nucleus. Indeed it was shown by $^{17}$O
NMR {} \cite{Yakubovskii_PRB67} that in
Pr$_{0.5}$Ca$_{0.5}$MnO$_{3}$ the CE-type magnetic correlations
develop gradually below $T_{CO}$. The AF correlations between the
$ab$-layers appear first, and only at a lower temperature the
CE-correlations in the $ab$-planes are formed. In this work we
compare the formation of the Mn-Mn spin correlations in Bi-based
manganites to that in Pr$_{0.5}$Ca$_{0.5}$MnO$_{3}$.

For half-doped manganites two possible scenario of
CO/OO-transition are considered. The first relates to the
localization of holes at $e_g$-orbital of the transition metal
cation. It results in the conventional CE-order of the ions of
different valence (Mn$^{3.5+\delta}$/Mn$^{3.5-\delta}$) in the
$ab$-plane. \cite{Wollan-Koeler_PR100,Jirak_JMMM53} This scenario
is based on an ionic description of the charge ordering, it is
also named site-centered. \cite{Efremov}

The second scenario suggests the formation of the Zener-polaron
state, which consists of FM-correlated dimers of Mn$^{3.5+}$-ions.
\cite{Daoud-Aladine_PRL89}

This option is also named bond centered. \cite{Efremov} The
relation between the kinetic energy of the electrons in the
$e_g$-band and the $ab$-plane superexchange coupling of
$t_{g}$-electrons defines, in many respects, which type of CO is
formed in a crystal. \cite{Efremov,Efremov_Nat_04} The details of
CO in manganites are still an open question. Nevertheless, it was
suggested in the $x$-ray diffraction study \cite{Hejtmanek_JAP93}
performed on a single crystal of Bi$_{0.5}$Sr$_{0.5}$MnO$_{3}$
 that the CO-state in this system is formed
by Zener pairs. We address the questions concerning the detailed
form of the CO phase by studying the relative intensity of the
lines in the $^{17}$O spectra measured for the completely formed
AF-CO state of  manganites.

\section{EXPERIMENT}

We used powder samples of Bi$_{0.5}$Ca$_{0.5}$MnO$_{3}$ ($T_{CO}$
= 330 K) \cite{Garcia_PRB63} and Bi$_{0.5}$Sr$_{0.5}$MnO$_{3}$
($T_{CO}$ = 475 K), \cite{Garcia_PRB63} which were prepared by the
standard solid state reaction technique. The powders were enriched
in $^{17}$O isotope up to $\approx 20 \% $. The single-phase
nature of the enriched samples was confirmed by $x$-ray
diffraction at room temperature.

The NMR measurements were performed with several commercial and
home-built pulse phase-coherent NMR spectrometers operated in the
frequency range 20 - 500 MHz using a spin-echo technique. The
$^{209}$Bi and $^{17}$O NMR spectra were obtained by measuring at
each frequency ($\nu$) the intensity, $^{209}$Int($\nu$), of the
spin-echo signal formed with the radio-frequency ($rf$) pulse
sequence $(\pi/2)-t-(\pi)$. The width of a ($\pi/2$) $rf$-pulse
did not exceed (1.5 - 2) $\mu$s. The $^{17}$O NMR signal in H$_{2}$O
was used as the frequency reference ($\nu_{0}$) to determine the
shift of NMR line $K = (\nu -\nu_{0})/\nu_{0}$.

The measurements of the broad $^{17}$O NMR spectra in the
antiferromagnetic state (see Section D) were performed with a
special care. We used a single NMR coil for the whole frequency
range. Its quality factor Q was chosen smoothly dependent on the
frequency $\nu$. At each frequency the impedance of the resonance
circuit was matched to the 50 Ohm input impedance of the receiver,
and the gain of the $rf$ power amplifier was adjusted to the
maximum of the echo intensity, $^{17}$Int, keeping the width of
the ($\pi/2$) $rf$-pulse constant. Such a care had to be paid in
order to provide at every point of the broad spectrum the
echo-signal voltage, $V(t = 0)$, which is proportional to spectral
intensity the $^{17}$O NMR spectrum multiplied by $\nu^{2}$. The
spectral intensity normalized in such a manner yields the fraction
of oxygen contributing to the echo at the given frequency point.

\bigskip

\section{ RESULTS AND DISCUSSION}

\subsection{ $^{209}$Bi zero-field NMR spectra}

\begin{figure}[!h]
\centerline{\includegraphics[width=0.95\hsize]{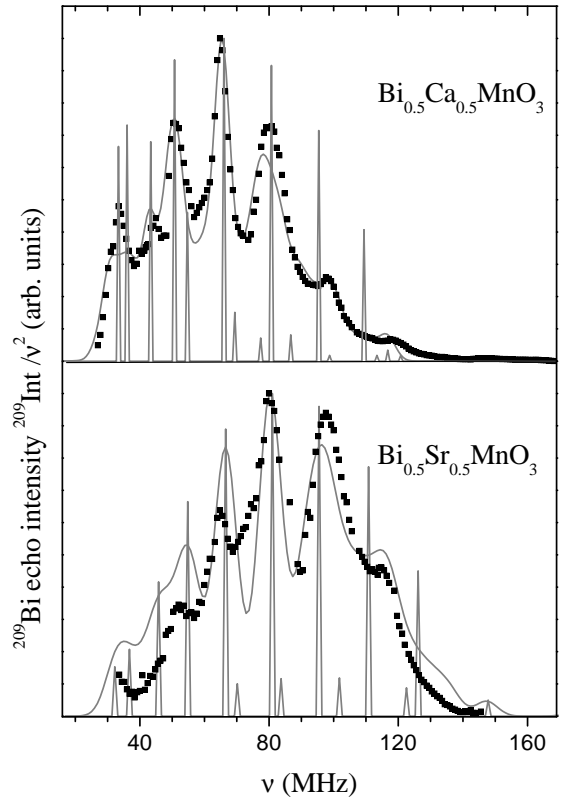}}
\caption{\label{fig_2} Zero-field $^{209}$Bi NMR spectra in
Bi$_{0.5}$Ca$_{0.5}$MnO$_3$ and Bi$_{0.5}$Sr$_{0.5}$MnO$_3$
measured by spin-echo technique at 4.2 K. The echo intensity
$^{209}$Int (solid squares) is plotted versus the frequency $\nu$.
The set of narrow peaks shows the calculated transitions with the
parameters given in Table~\ref{tab_1}. The grey curve is the
calculated broadened spectrum assuming a gaussian distribution for
both the local field $^{209}H_{loc}$ and  the quadrupole frequency
$\nu_Q$ (see text for details).}
\end{figure}

Figure~\ref{fig_2} shows the $^{209}$Bi NMR spectra measured in
the AFM-ordered state of Bi$_{0.5}$Ca$_{0.5}$MnO$_{3}$ ($T_{N} =
140$ K) \cite{Bokov_PSS20} and Bi$_{0.5}$Sr$_{0.5}$MnO$_{3}$
($T_{N} = 155$  K) \cite{Hejtmanek_JAP93} in zero external
magnetic field (ZFNMR) at $T = 4.2$ K. The spectral intensity of
the very broad line of $^{209}$Bi (nuclear spin $^{209}I = 9/2$,
quadrupolar moment $^{209}Q = - 0.35$ barns) is presented as the
echo intensity, $^{209}$Int($\nu$), corrected by the factor $~
\nu^{-2}$. The components of the local magnetic field
($^{209}H_{loc}$) as well as the parameters of electric field
gradient (EFG, $V_{ii}$)  - quadrupole frequency, $^{209}\nu_{Q} =
e^{2}QV_{ZZ}/21$, and asymmetry parameter, $^{209}\eta =
(V_{XX}-V_{YY})/V_{ZZ}$ with $|V_{ZZ}|\geq|V_{XX}|\geq|V_{YY}|$, -
were determined from the exact diagonalization of the Hamiltonian,
including both the magnetic and the electric quadrupole
interactions of the $^{209}$Bi nuclei with the corresponding local
magnetic field and EFG. The calculation was performed for the
matrix representation in the frame of reference, XYZ, where the
corresponding term of electric quadrupole interaction  is
diagonal. The calculated transitions  $\Delta m_{I} = 1; 2$ are
indicated by narrow peaks in Fig.~\ref{fig_2} and the
corresponding results for $^{209}$H$_{i}, ^{209}\nu_{Q}$ and
$^{209}\eta$ are listed in Table~\ref{tab_1}. We have taken into
account distributions of both $^{209}H_{loc}$ and the EFG
parameters, that can appear due to the Bi/Sr or Bi/Ca charge
disorder. The distribution functions were taken in the Gaussian
form with the corresponding width $\delta H_{loc}$ and
$\delta\nu_{Q}$. The calculated broadened spectrum representing
the result of such a convolution is drawn by the grey curve in
Fig.~\ref{fig_2}. With a single set of NMR parameters it
reproduces the experimental data points rather well in a wide
frequency range.
\begin{table}[!h]
\caption{\label{tab_1}Parameters of the calculated $^{209}$Bi
zero-field NMR spectra shown in Fig.~\ref{fig_1}.  See text for
the definition of the  X, Y and Z  components of $^{209}H_{loc}$.
$\delta H_{loc}$ and $\delta\nu_Q$  are the width of the gaussian
distribution of the magnetic and quadrupolar interactions,
respectively.}
\begin{ruledtabular}
\begin{tabular}{lcc}
Sample & Bi$_{0.5}$Sr$_{0.5}$MnO$_3$ & Bi$_{0.5}$Ca$_{0.5}$MnO$_3$
\\\hline $^{209}H_{loc}$ (kOe) & 101(5)
& 85(5) \\ $H_X$  (kOe) & 9 & 0 \\ $H_Y$  (kOe) & 37 & 33 \\
$H_Z$ (kOe) & 94 & 78 \\ $\delta H_{loc}$ (kOe) & 10 & 6 \\ $\nu_Q$  (MHz) & 16(1) & 14(1) \\
$\delta \nu_Q$   (MHz) & 3 & 3 \\ $\eta$ & 0.90(5) & 0.75(5)
\end{tabular}
\end{ruledtabular}
\end{table}

The line of central transition $m_{I} = -1/2 \longleftrightarrow
+1/2$ peaks at $ \nu_{0}$ = 80 MHz for
Bi$_{0.5}$Sr$_{0.5}$MnO$_{3}$ and at $\nu_{0}$ = 65 MHz for
Bi$_{0.5}$Ca$_{0.5}$MnO$_{3}$. It corresponds to a very large
local magnetic field at the Bi-site in the AFM-state
$^{209}H_{loc}$ = 101 kOe and $^{209}H_{loc}$ = 85 kOe in
Bi$_{0.5}$Sr$_{0.5}$MnO$_{3}$ and Bi$_{0.5}$Ca$_{0.5}$MnO$_{3}$,
respectively. For comparison, the local field at $^{139}$La nuclei
is $^{139}H_{loc}$ = 3.5 kOe in the AFM ordered LaMnO$_{3}$.
\cite{Kumagai_PRB59} Moreover, in the FM-ordered domains of the
half-doped La$_{0.5}$Ca$_{0.5}$MnO$_{3}$ the fully polarized eight
nearest Mn moments create at the $^{139}$La the field
$^{139}H_{loc}$ = 36.3 kOe \cite{Yoshinari_PRB60, Allodi_PRL60},
which is much less than those observed in this study.

Two types of magnetic order, namely A-type and CE$'$-type magnetic
structures were revealed at $T = 1.5$ K in neutron diffraction
studies of polycrystalline Bi$_{0.5}$Sr$_{0.5}$MnO$_{3}$.
\cite{Frontera_PRB64} The A-type magnetic order should result in
the complete cancellation of $^{209}H_{loc}$ at the Bi-site
similar to the situation of La-site in LaMnO$_{3}$.
\cite{Kumagai_PRB59} For CE$'$-type, the AFM-order leads to a net
transfer of spin polarization from the four neighboring
Mn$^{3+}$-ions to Bi (see Fig.~\ref{fig_1}~c).

The resulting local field is aligned along [001] direction and its
magnitude is proportional to $|\mu_{c}$(Mn$^{3+})|$- the
c-projection of the staggered magnetic moment of Mn$^{3+}$. For
further comparison of different hyperfine couplings it is more
convenient to consider the local field $^{209}h_{loc}$ created at
the Bi-nuclei by one electron spin. By specifying
$|\mu_{c}($Mn$^{3+})|$ = 1.4 $\mu_{B}$ \cite{Frontera_PRB64} we
obtain

\begin{equation}
^{209}h_{loc} = ^{209}H_{loc}/ [4 |\mu_{c}(Mn^{3+})|]= 18.2
\text{kOe}/\mu_{B} , \label{shift_1}
\end{equation}
for Bi$_{0.5}$Sr$_{0.5}$MnO$_{3}$ with the CE$'$-type AFM-order.
This value exceeds more than four times the corresponding value of
$^{139}h_{loc}$ = 4.5 kOe/$\mu_{B}$ found in
La$_{0.5}$Sr$_{0.5}$MnO$_{3}$ at the FM state.
\cite{Yoshinari_PRB60}

It is worth noting that in the insulating state of manganites the
$6sp$-orbital of La$^{3+}$-ion is nearly empty, while the
Bi$^{+3}$-ion holds two electrons at the $6sp$ valence shell. As
it was suggested in Ref.~\onlinecite{Yoshinari_PRB60}, the
$6s$-orbital of the La$^{3+}$ cation can hybridize with one of the
three $t_{2g}$(Mn) atomic orbitals with a single unpaired spin. In
the Bi-based manganites the similar $t_{2g}$ (Mn)-$6s$(Bi) overlap
is one of the possible ways to transfer $3d$-electron spin
polarization of Mn-ions to the central Bi-cation. Another
possibility is the Mn spin polarization transfer via
$e_{g}$(Mn)-$2p$(O)-$6p$(Bi) orbitals but it seems less effective
in providing a local field comparable in magnitude with the
observed one. Indeed, the corresponding hyperfine field due to the
core polarization $H_{cp}(6p)\sim 300$ kOe per electron is
approximately two orders of magnitude less than the Fermi contact
hyperfine field $H_{FC}(6s)\sim 10^{4}$ kOe per electron.
\cite{Carter_MS20} The former case would lead to an unrealistic
high value of the $p$-wave spin polarization $f_{6p}$, which is
transferred from Mn to the $6p$-orbital of Bi - $f_{6p} =$$
^{209}h_{loc}/ H_{cp}(6p) \thickapprox 0.15$. In the case of
$t_{2g}$ (Mn)-$6s$(Bi) overlap, the large Fermi-contact field
$H_{FC}(6s)$ yields a much more reasonable estimate of the
corresponding transferred $6s$-wave spin polarization $f_{6s}=$$
^{209}h_{loc}/H_{FC}(6s) \thickapprox 0.004$.

Thus, taking into account the large Fermi-contact hyperfine field
created at the nuclei by a single unpaired electron located at the
$6s$-orbital, $H_{FC}(6s)$, one can suggest that the observed very
large $^{209}H_{loc}$ strongly favors the dominant $6s$ character
rather than the $p$ character of the lone electron pair of Bi in
both Bi$_{0.5}$Sr$_{0.5}$MnO$_{3}$ and
Bi$_{0.5}$Ca$_{0.5}$MnO$_{3}$.

We do not think that the reduced value of $^{209}H_{loc}$ in
Bi$_{0.5}$Ca$_{0.5}$MnO$_{3}$ compared to
Bi$_{0.5}$Sr$_{0.5}$MnO$_{3}$ is due to a change in
$s^{2}\longrightarrow sp/p^{2}$ character of the Bi lone electron
pair as suggested in Ref.~\onlinecite{Garcia_PRB63}, since the
decrease is not as large as expected for such change when
considering the values of $H_{FC}(6s)$ and $H_{cp}(6p)$.

The difference in the observed $^{209}H_{loc}$ is more probably
due to variation of the canting of the staggered magnetic moments
of Mn$^{3+}$ ions in Bi$_{0.5}$Ca$_{0.5}$MnO$_{3}$ compared with
Bi$_{0.5}$Sr$_{0.5}$MnO$_{3}$ as shown below. As the magnetic
structure of the AFM state in Bi$_{0.5}$Ca$_{0.5}$MnO$_{3}$ at low
temperature has not been reported yet we make an attempt to
clarify this point by measuring $^{17}$O ZFNMR in the same samples
at 4.2K.

\subsection{ $^{17}$O Zero Field NMR spectra}

\begin{figure}[h]
\centerline{\includegraphics[width=0.95\hsize]{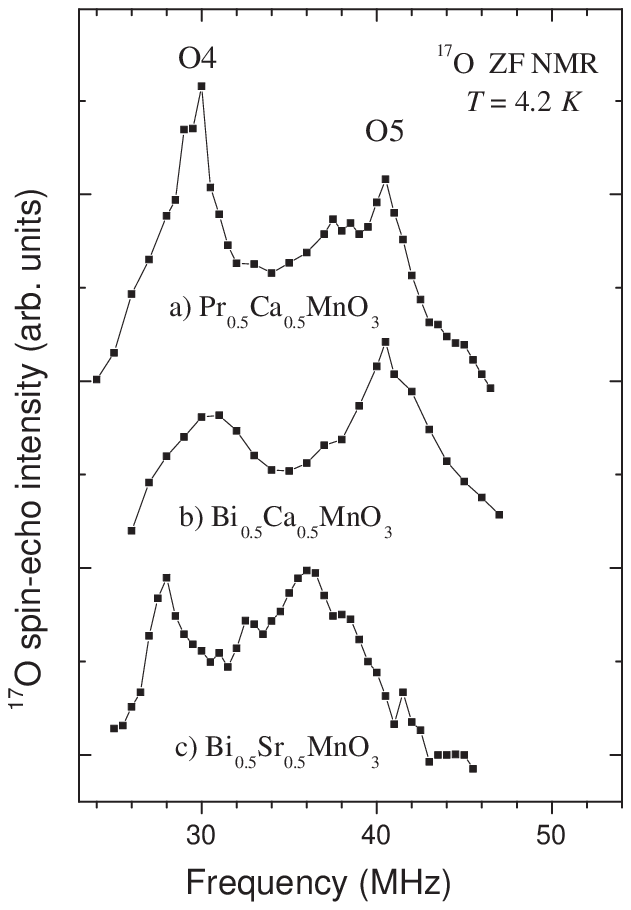}}
\caption{\label{fig_3} Zero-field $^{17}$O NMR spectra measured at
4.2 K in the antiferromagnetic charge ordered state of
Pr$_{0.5}$Ca$_{0.5}$MnO$_3$ (a) Bi$_{0.5}$Ca$_{0.5}$MnO$_3$ (b)
and Bi$_{0.5}$Sr$_{0.5}$MnO$_3$ (c). The solid lines are a guide
for the eye.}
\end{figure}

Figure~\ref{fig_3} shows the zero field $^{17}$O NMR (ZFNMR)
spectra in the AFM ordered state of Pr$_{0.5}$Ca$_{0.5}$MnO$_{3}$
(a), Bi$_{0.5}$Ca$_{0.5}$MnO$_{3}$ (b), and
Bi$_{0.5}$Sr$_{0.5}$MnO$_{3}$ (c) oxides at 4.2 K. \footnote{In
both Bi-based manganites the ZFNMR spectrum shows a partial
overlap of the low-frequency tail of the $^{209}$Bi signal with
the $^{17}$O spectrum (see Fig.~\ref{fig_2} and \ref{fig_3}). The
spin-echo spectra of $^{17}$O (Fig.~\ref{fig_3}) were obtained by
applying a pulse sequence, which parameters were adjusted to
obtain the maximum ratio I = $^{17}$Int/$^{209}$Int where
$^{17}$Int and $^{209}$Int are the spin echo intensities. In these
conditions the ratio $I$ exceeds 10. Furthermore, we have compared
the ZFNMR spectra measured in the $^{17}$O enriched and in a
natural Bi$_{0.5}$Ca$_{0.5}$MnO$_{3}$ sample. The latter is a
non-enriched fraction of the very same
Bi$_{0.5}$Ca$_{0.5}$MnO$_{3}$ material. Its $^{209}$Bi ZFNMR
spectrum is shown on the top panel of Fig.2. From this comparison,
we found that the difference of the spectra reproduces rather well
the two-peaks line shape and each peak position of the $^{17}$O
ZFNMR spectrum presented on Fig.~\ref{fig_3}~b.}

\begin{table*}[]
\caption{\label{tab_2}Zero-field $^{17}$O NMR resonance
frequencies for two oxygens sites in the $ab$-plane of
Bi$_{0.5}$Sr$_{0.5}$MnO$_3$ (CE$'$-type magnetic structure),
Bi$_{0.5}$Ca$_{0.5}$MnO$_3$ and Pr$_{0.5}$Ca$_{0.5}$MnO$_3$
(CE-type magnetic structure) at 4.2 K. The corresponding estimated
values of $^{17}H_{loc} = 2\pi\nu/^{17}\gamma$ are shown in
slashes. For Bi$_{0.5}$Sr$_{0.5}$MnO$_3$ the components of the
magnetic moment $\mu_{ab}$ and $\mu_c$ of Mn$^{3+}$/Mn$^{4+}$-ions
at 4.2 K as well as $\beta$, the angle between $\mu$(Mn$^{3+}$)
and the $ab$ plane, were calculated using the $^{209}$Bi and
$^{17}$O ZFNMR data (see text for details). For
Bi$_{0.5}$Sr$_{0.5}$MnO$_3$ and Pr$_{0.5}$Ca$_{0.5}$MnO$_3$ these
parameters were measured in Ref.~\onlinecite{Frontera_PRB64} and
Ref.~\onlinecite{Jirak_PRB61} respectively.}

\begin{ruledtabular}
\begin{tabular}{rcccc}
\multicolumn{2}{c}{Sample} & Bi$_{0.5}$Sr$_{0.5}$MnO$_3$ &
Bi$_{0.5}$Ca$_{0.5}$MnO$_3$ & Pr$_{0.5}$Ca$_{0.5}$MnO$_3$ \\\hline
\multicolumn{2}{c}{$\nu$(O4) (MHz)/$^{17}H_{loc}$(O4) (kOe)/} & 28/48.5/  & 31/54/ & 30/52/ \\
\multicolumn{2}{c}{$\nu$(O4) (MHz)/$^{17}H_{loc}$(O4) (kOe)/} & 36/62/ & 40/69/ & 40/69/ \\
\multicolumn{1}{r}{$\mu$ ($\mu_B$)}& $ \left\{ \begin{tabular}{l} $|\mu_{ab}$(Mn$^{4+}$)$|$\\
$|\mu_{c}$(Mn$^{4+}$)$|$\\
$|\mu_{ab}$(Mn$^{3+}$)$|$\\
$|\mu_{c}$(Mn$^{3+}$)$|$\end{tabular} \right. $ &
\begin{tabular}{c}
1.6\\
1.6\\
2.1\\
1.4\end{tabular} &
\begin{tabular}{c}
1.8\\

1.8\\
2.5\\
1.15\end{tabular} &
\begin{tabular}{c}
2.75\\
0\\
3.18\\
0\end{tabular} \\
\multicolumn{2}{c}{$\beta$   ($^\circ$)}& 33 & 25 & 0
\end{tabular}
\end{ruledtabular}
\end{table*}

All three spectra consist of two rather broad peaks, the positions
of their maxima are listed in Table \ref{tab_2}. For proper
attribution of the peaks to each of the (O1)-(O5)
\cite{Jirak_PRB61} oxygen sites in the Bi-based manganites we
first focus on Pr$_{0.5}$Ca$_{0.5}$MnO$_{3}$ which has the
canonical CE-type magnetic structure at low temperature.

The perfect CE-type CO/OO implies a zigzag arrangement of the
ordered $e_{g}(3x^{2}-r^{2})$ and $e_{g}(3y^{2}-r^{2})$ orbitals
of Mn$^{3+}$-ions in the $ab$-plane of the orthorhombic $Pbnm$
lattice, with the FM correlated magnetic moments in the zigzags.
The neighboring zigzags are AF-coupled, the ordering in
$c$-direction is also AF. Oxygen in the apical positions of
MnO$_{6}$ octahedra is located between two Mn$^{3+}$-ions
(O1-site, Fig.~\ref{fig_3} in Ref.~\onlinecite{Jirak_PRB61}) or
between two Mn$^{4+}$-ions (O2/O3-site). As shown in
Ref.~\onlinecite{Yakubovskii_PRB67}, the local magnetic field
$^{17}H_{loc}$ is created mainly by the $2s$ spin polarization of
oxygen transferred via the $e_{g}$-(Mn)-$2s$(O) overlap from the
neighboring Mn. This spin arrangement causes complete cancellation
of the local magnetic field at the apical O1-O3 sites, and its
corresponding ZFNMR line is out of the frequency range presented
in Fig.~\ref{fig_3}. The O4-O5 sites correspond to the oxygens in
the $ab$-plane. The oxygen (O4-site) participates in the
AF-coupling of neighboring Mn$^{4+}$, Mn$^{3+}$ from adjacent
zigzags. The other oxygen (O5-site) is located inside a zigzag
surrounded by two FM-coupled Mn$^{4+}$ and Mn$^{3+}$ ions.

The high-frequency peak in the $^{17}$O ZFNMR spectrum is
attributed to oxygen positioned in O5-sites whereas the
low-frequency peak is presumably due to oxygen located in
O4-sites. Indeed for oxygen in O5-site the transferred $s$-wave
spin density is maximal since within the zigzag the lobe of the
partially occupied $e_{g} (m_{l}=0)$ orbital of Mn$^{3+}$-ion
points toward this oxygen. Furthermore two neighboring Mn-ions are
FM-correlated. For O4-site a rather large transferred hyperfine
field is expected for the following reason. Although the Mn spins
from adjacent zigzags are antiparallel, the O4-oxygen is
"sandwiched" between Mn$^{3+}$ and Mn$^{4+}$-ions with different
spin values and different orbital occupations, i.e. with different
covalency. So the transferred polarizations from these two Mn-ions
should not compensate each other  as they do for the apical
oxygens (O1,O2/O3). It results in a substantial but smaller shift
than for the O5-site. Moreover the transferred $s$-wave
polarization from Mn$^{4+}$-ion is expected to be negative due to
effects of covalent mixing with the empty $e_{g}$-orbitals.
Indeed, the charge transfer from the occupied O-$2s$ orbital to
the empty $e_g$-orbital is spin-dependent since it is regulated by
the intra-atomic exchange coupling with electrons on
$t_{g}$-orbitals.

In the CE spin ordered state the transferred static s-wave
polarization at the oxygen site may be expressed through the
transferred spin densities $f_{s,3+}\mu $(Mn$^{3+}$) or
$f_{s,4+}\mu$(Mn$^{4+}$) of the neighboring Mn (with $f_{s,3+} > 0
$ and $f_{s,4+} < 0$), and the following expressions can be
considered for $^{17}H_{loc}$ at the O5-site:

\begin{eqnarray}
^{17}H_{loc} (O5) = (^{17}\gamma
\hbar)^{-1}a(2s)\{f_{s,3+}\mu(Mn^{3+}) + \nonumber \\
f_{s,4+}\mu(Mn^{4+})\},\label{shift_2}
\end{eqnarray}
and at the O4-site:

\begin{eqnarray}
^{17}H_{loc} (O4) = (^{17}\gamma
\hbar)^{-1}a(2s)\{0.25f_{s,3+}\mu(Mn^{3+}) -\nonumber
\\ f_{s,4+}\mu(Mn^{4+})\}, \label{shift_3}
\end{eqnarray}
respectively. $H_{loc}$, $\mu$(Mn$^{3+}$) and $\mu$(Mn$^{4+}$) are
vectors. The sign "+"/"-" in \ref{shift_2}, \ref{shift_3} refers
to the FM/AF spin correlations of the neighbor Mn$^{3+}$- and
Mn$^{4+}$-ions. Here, $a(2s) =$$ ^{17}\gamma\hbar
H_{FC}(2s)$=0.0216 cm$^{-1}$ is the isotropic hyperfine coupling
constant for oxygen ion. \cite{Fraga_Handbook} $H_{FC}(2s)$ = 1.1
MOe is the corresponding hyperfine magnetic field due to the
Fermi-contact interaction with the electron located on the
$2s$-orbital. Following Ref.~\onlinecite{Shulman_PR108}, the
corresponding isotropic spin density transferred to oxygen from
neighboring Mn-ions may be defined with the factor $f_{s} =
H_{loc}/2H_{FC}(2s)$.

Using the $^{17}$O ZFNMR results and the neutron diffraction data
from Ref.~\onlinecite{Jirak_PRB61}, listed in Table~\ref{tab_2},
for Pr$_{0.5}$Ca$_{0.5}$MnO$_{3}$  we can quantify the factors
$f_{s}$ in Eqs.~(\ref{shift_2}),~(\ref{shift_3}) as: $f_{s,3+} =
0.0075$ and $f_{s,4+}= - 0.0025$, which are somewhat less than
$f_{s} = 0.01$ (Ref.~\onlinecite{Yakubovskii_PRB67}) estimated in
a similar way in the CD PM state for the very same sample.

Let us now consider Bi$_{0.5}$Sr$_{0.5}$MnO$_{3}$. As the degree
of the spatial overlap of $e_g$(Mn)-$2_s$(O) orbital decreases
rather slowly with the interatomic Mn-O distance in the series of
manganites considered here, we assume that the ratio
$f_{s,3+}/|f_{s,4+}|$ is not changed from
Pr$_{0.5}$Ca$_{0.5}$MnO$_{3}$ to Bi-manganites. With this
assumption and using the value of $\mu$ deduced from the refined
neutron diffraction data for CE$'$-phase in
Bi$_{0.5}$Sr$_{0.5}$MnO$_{3}$, \cite{Frontera_PRB64} equations
(\ref{shift_2}) and (\ref{shift_3}) yield the values $f_{s,3+} =
0.009$ and $f_{s,4+}= - 0.003$. Thus one can conclude that the
admixture of the $s$-wave character to the $e_g$-wave function
slightly increases in Bi$_{0.5}$Sr$_{0.5}$MnO$_{3}$ compared to
Pr$_{0.5}$Ca$_{0.5}$MnO$_{3}$

For Bi$_{0.5}$Ca$_{0.5}$MnO$_{3}$, there are no available refined
staggered magnetization data. We have used the following
assumptions in order to calculate the magnetic moments with the
help of Eqs.(\ref{shift_1})-(\ref{shift_3}):

- the same CE$'$-type magnetic structure as in
Bi$_{0.5}$Sr$_{0.5}$MnO$_{3}$ takes place in
Bi$_{0.5}$Ca$_{0.5}$MnO$_{3}$ at low temperature.

- the $s$-wave admixture to a character of the lone electron pair
of the Bi$^{3+}$-ion, $f_{6s}$, is kept constant from Sr to Ca
manganite.

As can be seen in Table \ref{tab_2}, from BiSrMnO to BiCaMnO to
PrCaMnO the magnitude $|\mu|$ increases. Furthermore, it is
accompanied by a decrease of $\mu_c$ which indicates a decreasing
canting. Compared to the CE structure of
Pr$_{0.5}$Ca$_{0.5}$MnO$_{3}$, the CE$'$ structure of the Bi-based
samples may be related to a stronger interplane coupling and/or a
specific anisotropy of the Jahn-Teller distortion along $c$ axis.
Our results show that the CE$'$-type structure in
Bi$_{0.5}$Sr$_{0.5}$MnO$_{3}$  is more pronounced than in
Bi$_{0.5}$Ca$_{0.5}$MnO$_{3}$.

\subsection{ $^{17}$O NMR in the paramagnetic charge-disordered and
charge-ordered states}

\begin{figure*}
\centerline{\includegraphics[width=0.95\hsize]{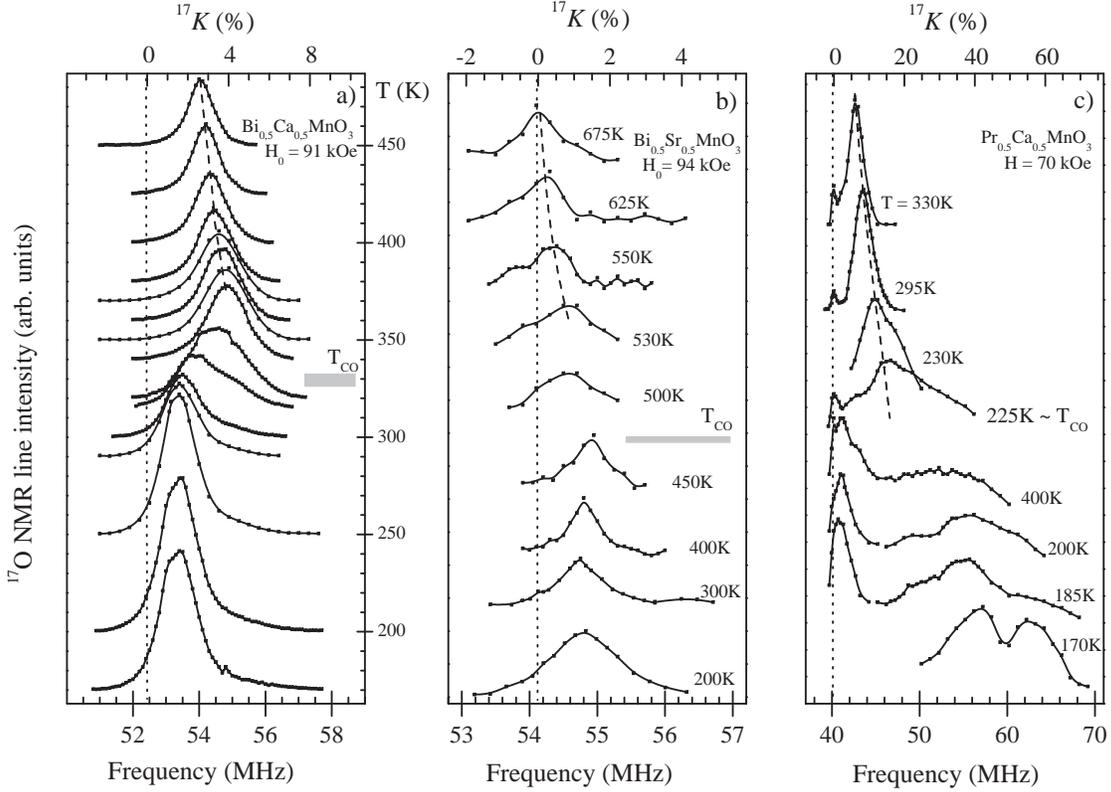}}
\caption{\label{fig_4} $^{17}$O NMR spectra measured in the
paramagnetic state of Bi$_{0.5}$Ca$_{0.5}$MnO$_3$ (a),
Bi$_{0.5}$Sr$_{0.5}$MnO$_3$ (b), and Pr$_{0.5}$Ca$_{0.5}$MnO$_3$
(c). \cite{Yakubovskii_PRB67} The position of the Larmor frequency
($K$ = 0) is shown by dotted vertical line. The grey narrow stripe
indicates $T_{CO}$ - critical temperature of the charge ordering.}
\end{figure*}

Figures~\ref{fig_4}~(a,b) show the $^{17}$O NMR spectra measured
at different temperatures  on Bi$_{0.5}$Ca$_{0.5}$MnO$_{3}$ and
Bi$_{0.5}$Sr$_{0.5}$MnO$_{3}$ in the PM state. The spectra were
obtained at $H \thickapprox$ 90 kOe above $T=320$ K and at $H
\thickapprox$ 120~kOe below 320 K. The comparison (not shown) of
the spectra obtained with both fields around 300K demonstrates the
negligible effect of the quadrupolar interaction. For comparison
we reproduce in Fig.~\ref{fig_4}~(c) the $^{17}$O NMR spectra ($H
\thickapprox$ 70 kOe) of Pr$_{0.5}$Ca$_{0.5}$MnO$_{3}$ in the PM
state. \cite{Yakubovskii_PRB67} All the spectra were taken when
cooling the samples from the highest temperature indicated in
Fig.~\ref{fig_4}~(a-c) to avoid any hysteresis. According to
resistivity data, \cite{Tomioka_PRB53, Kirste_PRB67} cooling our
Bi-samples in the external magnetic field indicated above does not
suppress the CO-transition.

For all compounds, in the CD-state above $T_{CO}$ the $^{17}$O
spectrum consists of a single line. With decreasing temperature
its shift $^{17}K$ increases following the Curie-Weiss law $^{17}K
\sim (T - \theta)^{-1}$ with positive $\theta$ indicating the
presence of FM electron correlations of Mn: $\theta = 250(40)$~K
in Bi$_{0.5}$Sr$_{0.5}$MnO$_{3}$, 170(10) K in
Bi$_{0.5}$Ca$_{0.5}$MnO$_{3}$ and 130(20) K in
Pr$_{0.5}$Ca$_{0.5}$MnO$_{3}$. \cite{Yakubovskii_PRB67}

We have used static magnetic susceptibility $\chi$ data for
Bi$_{0.5}$Ca$_{0.5}$MnO$_{3}$ \cite{Bokov_PSS20} and
Bi$_{0.5}$Sr$_{0.5}$MnO$_{3}$ \cite{Hejtmanek_JAP93} to estimate
the corresponding $s$-wave spin density transferred to oxygen from
neighboring Mn in the CD state. The slope of $^{17}K$ $vs$ $\chi$
plot corresponds to the local magnetic field $^{17}H_{loc} =
\mu_{B}\Delta K/\Delta\chi$ =12.0(5) kOe/$\mu_{B}$
(Bi$_{0.5}$Sr$_{0.5}$MnO$_{3}$) and $^{17}H_{loc}$ = 12.4(5)
kOe/$\mu_{B}$ (Bi$_{0.5}$Ca$_{0.5}$MnO$_{3}$ ). For
Pr$_{0.5}$Ca$_{0.5}$MnO$_{3}$ we refined the estimation of
$^{17}H_{loc}$ = 7.1(3) kOe/$\mu_{B}$ in an accurate $^{17}$O NMR
study performed at 120 kOe as compared to the earlier published
value. \cite{Yakubovskii_PRB67} In comparison to
Pr$_{0.5}$Ca$_{0.5}$MnO$_{3}$,  the increase of $^{17}H_{loc}$ in
the Bi-based manganites evidences a larger $e_{g}$(Mn)-$2s$(O)
admixture.

For Pr$_{0.5}$Ca$_{0.5}$MnO$_{3}$, $T_{crit} = 250$ K.
\cite{Jirak_PRB61} $T_{crit}$ is defined as the temperature at
which the transition from the high-$T$ cubic $Pm3m$ to the
orthorhombic $Pbnm$ structure occurs and thus it is related to the
onset of orbital ordering. As seen in Fig.~\ref{fig_4}~(c), below
$T \thickapprox 250$ K the $^{17}$O spectrum of
Pr$_{0.5}$Ca$_{0.5}$MnO$_{3}$ substantially broadens and splits
into three lines as the temperature approaches $T_{N}$. According
to Ref.~\onlinecite{Yakubovskii_PRB67}, the low-frequency line is
attributed to apical oxygens while the two other lines
corresponding to the shift $^{17}K=40\%$ and $55\%$ are attributed
to the oxygens located in $ab$-plane (O4- and O5-site,
respectively). As said before, the magnetic shift $^{17}K$ of the
apical line is greatly reduced due to the vanish of the local
field. The NMR signal of the apical oxygen appears at a
characteristic temperature $T_{CO,nmr}$, below which AF spin
correlations between $ab$-layers occur. It is not surprising that
$T_{CO,nmr}$ coincides with $T_{crit}$ since the localization of
the itinerant electrons at $e_{g}$-orbitals should manifest itself
both in the appearance of AF spin correlations between $ab$-layers
and in the cooperative Jahn-Teller distortions in the
MnO$_{6}$-octahedra sublattice.

\begin{figure}[h]
\centerline{\includegraphics[width=0.95\hsize]{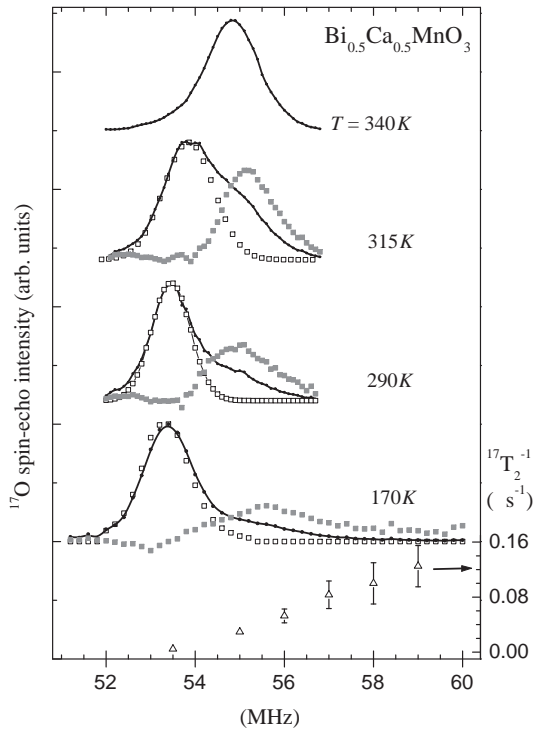}}
\caption{\label{fig_5} Evolution of the $^{17}$O spectrum pattern
for Bi$_{0.5}$Ca$_{0.5}$MnO$_3$ showing the onset of the splitting
into two lines when cooling below $T_{CO}$ from the
charge-disordered to the charge- ordered paramagnetic phase. The
solid line shows the echo intensity vs frequency measured with the
time delay $t$ = 12 $\mu$s between the $rf$-exciting pulses; open
squares indicate the echo intensity measured with $t$ = 100
$\mu$s. Solid squares are the normalized high-frequency part of
the spectra obtained by reconstructing at $t$ = 0 the difference
between the echo intensities measured with 12 $\mu$s and 100
$\mu$s. The open up-triangles are the $^{17}T_{2}^{-1}$ data
measured between 53 MHz and 59 MHz at $T = 170$ K.}
\end{figure}

For Bi$_{0.5}$Ca$_{0.5}$MnO$_{3}$, the $^{17}$O NMR spectra
drastically change below $T_{CO,nmr}$, indicated by the narrow
grey stripe in Fig.~\ref{fig_4}~(a). Moreover, the line shape
shown in Fig.~\ref{fig_5} becomes extremely sensitive to $t$ - the
time delay between the $rf$-exciting pulses. The solid curve shows
the echo intensity $vs$ frequency ($\nu$) measured at short time
delay $t$ = 12 $\mu$s, while the open squares indicate the echo
intensity observed at large delay, $t$=100 $\mu$s. By measuring
the rate of the echo decay ($^{17}T_{2}^{-1}$) at several
representative points of the spectrum, we have reconstructed the
high-frequency part of the spin-echo spectrum at $t$ = 0 (solid
squares), its intensity being the difference between the spectra
measured with $t$ = 12 $\mu$s and 100 $\mu$s. It worth noting
that, $T_{CO,nmr}$ (Fig. ~\ref{fig_5}) is consistent with
$T_{crit}=335(30)$ K (Ref.~\onlinecite{Bokov_PSS20}) defined as
the temperature of the cubic-to-orthorhombic structural
transition. With further decrease of temperature a substantial
part of the high-frequency tail becomes undetectable due to a too
small value of $T_{2}$ ($^{17}T_{2, min} \thickapprox$ 5 $\mu$s).
Although only Bi$_{0.5}$Ca$_{0.5}$MnO$_{3}$ is shown in
Fig.~\ref{fig_5}, a very similar $^{17}T_{2}$ effect on the
high-frequency tail of the $^{17}$O spin-echo spectrum occurs in
the CO-phase of Bi$_{0.5}$Sr$_{0.5}$MnO$_{3}$ below $T_{CO,nmr}$.

\begin{figure}[h]
\centerline{\includegraphics[width=0.95\hsize]{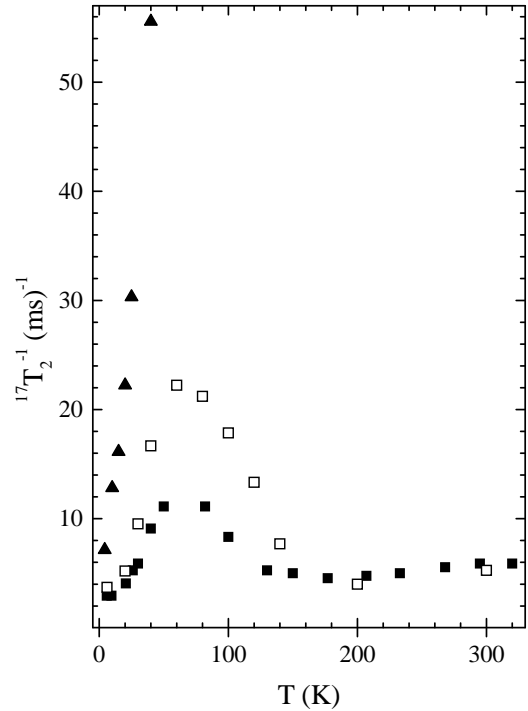}}
\caption{\label{fig_6} Thermal behaviour of $^{17}$O spin-spin
relaxation rate, $^{17}T_{2}^{-1}$, in the charge ordered,
paramagnetic and antiferromagnetic phases for the two Bi-based
manganites. $^{17}T_{2}^{-1}$ of the oxygen in the $ab$-plane
(O4-site) measured with $H_{ext} = 0 - $($\vartriangle$) and of
the apical oxygen ($\blacksquare$) measured with $H_{ext}$ = 50
kOe in Bi$_{0.5}$Ca$_{0.5}$MnO$_3$. For
Bi$_{0.5}$Sr$_{0.5}$MnO$_3$, $^{17}T_{2}^{-1}$ of the apical
oxygen ($H_{ext} = 120$ kOe) is shown by open squares.}
\end{figure}

In order to understand why in contrast to
Pr$_{0.5}$Ca$_{0.5}$MnO$_{3}$ below $T_{CO,nmr}$, the
high-frequency part of the spectrum in
Bi$_{0.5}$Ca$_{0.5}$MnO$_{3}$ and Bi$_{0.5}$Sr$_{0.5}$MnO$_{3}$ is
difficult to be detected, we have studied $^{17}T_{2}$ thermal
behaviour in both CO phases, paramagnetic and antiferromagnetic.
$^{17}T_{2}^{-1}$, measured in Bi$_{0.5}$Ca$_{0.5}$MnO$_{3}$ on
the low frequency peak of the ZF NMR spectrum (O4 site, $\nu$ = 30
MHz; see Fig.~\ref{fig_3}~b) is shown by up-triangles in
Fig.~\ref{fig_6}. It evidences a divergent increase with
temperature. The data set shown by the closed and open squares
correspond to $^{17}T_{2}^{-1}$ measured on the apical line in
Bi$_{0.5}$Sr$_{0.5}$MnO$_{3}$ and in Bi$_{0.5}$Ca$_{0.5}$MnO$_{3}$
respectively. For the apical line there is also an enhancement of
$^{17}T_{2}^{-1}$ which occurs in the same temperature range for
both Bi-based compounds. Furthermore, though weaker, this
enhancement occurs in the same temperature range as for O4 site.
The detailed study of the Mn-spin fluctuation probed by
$^{17}T_{2}^{-1}$ will be discussed elsewhere. Here we just point
out the $T_2$ which is more than one order of magnitude shorter
than in Pr$_{0.5}$Ca$_{0.5}$MnO$_{3}$ and thus the difference in
the Mn spin fluctuation in the Bi-based manganites compared to
Pr$_{0.5}$Ca$_{0.5}$MnO$_{3}$.

The thermal behaviour of $^{17}T_{2}^{-1}$  explains why in the
Bi-based compounds the high frequency lines corresponding to O4
and O5, which are not detected in the PM CO phase
($^{17}T_{2}^{-1}$ $>$ 100 ms$^{-1}$), are well observed at very
low temperature. Indeed, they appear at $T$ = 4.2 K as a
two-peaked spectrum in zero field NMR
(Figs.~\ref{fig_3}~b;~\ref{fig_3}~c) and as a very broad
high-frequency tail at $T$ = 15 K for the $^{17}$O NMR spectrum
measured at 12 T as shown in Fig.~\ref{fig_7} for
Bi$_{0.5}$Sr$_{0.5}$MnO$_{3}$.

\subsection{ $^{17}$O NMR in the antiferromagnetic charge-ordered
state}

In this section we consider the peculiarities in the $^{17}$O NMR
spectrum measured in the antiferromagnetic charge-ordered state of
Bi$_{0.5}$Sr$_{0.5}$MnO$_{3}$  far below the Neel temperature,
when the development of the static spin correlations is completed.
We show that our analysis of the spectrum yields some constrains
for the theoretical model of the CO phase in manganites.

\begin{figure}[!h]
\centerline{\includegraphics[width=0.95\hsize]{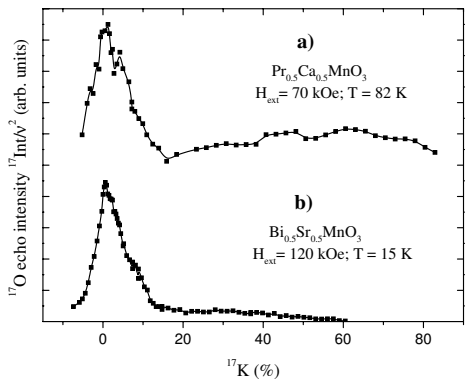}}
\caption{\label{fig_7} $^{17}$O NMR spectrum in the
antiferromagnetic charge-ordered phase of
Pr$_{0.5}$Ca$_{0.5}$MnO$_3$ (a) and Bi$_{0.5}$Ca$_{0.5}$MnO$_3$
(b). The spectra were obtained at the field cooled samples.}
\end{figure}

The $^{17}$O NMR spectra of Pr$_{0.5}$Ca$_{0.5}$MnO$_{3}$ and
Bi$_{0.5}$Sr$_{0.5}$MnO$_{3}$  are shown in Figure~\ref{fig_7}.
The intensity is a result of two corrections. First, the intensity
being related to the $T_{2}$ value at a given frequency, the
spectra were recalculated for a time delay $t$ = 0, between
exciting $rf$-pulses. Second, the spectral intensity is
proportional to the intensity of the echo signal, $^{17}$Int,
normalized by the factor $\nu^{-2}$ at each frequency point $\nu$
of the spectrum. Thus, it yields directly the fraction of oxygen
at (O1-O5) sites, which contributes to the echo-signal at a given
frequency point $^{17}K = (\nu - \nu_{0})/\nu_{0}$.

For Pr$_{0.5}$Ca$_{0.5}$MnO$_{3}$   the spectrum was measured in
the same magnetic field of 70 kOe as in the CO paramagnetic phase
shown in Fig.~\ref{fig_4}~c. The narrow line-A is due to oxygen in
the apical (O1-O3) sites, while the broad line-(B+C) originates
from oxygen in the $ab$-plane (O4-O5 sites). The unresolved
pattern of this line is a result of additional magnetic broadening
below $T_{N} = 170$ K. The ratio of the integral NMR line
intensities $r_{nmr} = (I_A/ I_{B+C} ) = 0.96(10)$ is very close
to the structural ratio $r_{struct}= 2:2$ of the oxygen atoms
occupying the apical sites and those placed in the $ab$-plane.

It is worth emphasizing  that this estimate of $r_{nmr}
\thickapprox r_{struct} =1$ allows the scenario of the
site-centered CE-type charge ordering which develops in
Pr$_{0.5}$Ca$_{0.5}$MnO$_{3}$ among the ions of different valence
(Mn$^{3.5+\delta}$/Mn$^{3.5-\delta}$).
\cite{Wollan-Koeler_PR100,Jirak_JMMM53,Efremov} Indeed, such a
value is expected only for the magnetic structure of the
conventional CE-type, i.e., in the ionic model.

Let us address now to the $^{17}$O NMR spectrum of
Bi$_{0.5}$Sr$_{0.5}$MnO$_{3}$ measured at 120 kOe and 15 K . The
essential feature of the spectrum is that the gravity center of
the broad high-frequency line is substantially shifted towards
smaller $^{17}K$, compared with the rather well resolved spectrum
of Pr$_{0.5}$Ca$_{0.5}$MnO$_{3}$. The integral intensity of line-A
is approximately twice as larger as that of the broad line
integrated in the range of $^{17}K$ above $15\%$. The
corresponding ratio $r_{nmr} \thickapprox 2$ is a clear indication
that in the CO state of Bi$_{0.5}$Sr$_{0.5}$MnO$_{3}$ a sizeable
fraction of oxygen located in the $ab$-plane experiences a reduced
local field. The reduction of this local field is mainly
controlled by the occupancy of the $e_{g}$(Mn) orbital. The
inequality $r_{nmr}> r_{struct}=1$ is expected for the
bond-centered models of CO suggested for half doped manganites.
\cite{Daoud-Aladine_PRL89,Efremov, Efremov_Nat_04} In these OO
models there is a reduction of $^{17}H_{loc}$ at some oxygen sites
in the $ab$-plane. As an example, for the Zener polaron state,
\cite{Daoud-Aladine_PRL89} the local field should vanish at the
oxygen atom located between the AF-correlated adjacent dimers of
Mn$^{3.5+}$. Thus, we incline to consider the $e_{g}$(Mn) orbital
ordering in the CO state of Bi$_{0.5}$Sr$_{0.5}$MnO$_{3}$ in the
frame of the CO bond-centered models. It's worth noting that for
the CE$'$-type magnetic structure, the Mn magnetic moments do not
lie in the $ab$-plane and the $c$-component $\mu_{c}\neq 0$.  In
the CE$'$-type structure, the conventional theoretical treatment
of the OO order in the frame of 2D-models is not appropriate
approach to consider in join the $e_{g}$-orbital and spin order,
observed in Bi$_{0.5}$Sr$_{0.5}$MnO$_{3}$ at low $T$.\footnote{For
the CE$'$ magnetic structure of the Mn$^{3+/4+}$ ions ordered in
the ab-plane like a checkerboard (see Fig.~\ref{fig_2}~a in
Ref.~\onlinecite{Frontera_PRB64}) one might expect $r_{nmr} < 1$,
since an apical oxygen surrounded by Mn$^{3+}$ ions (see
Fig.~\ref{fig_1}~c) experiences the local field not cancelled
along $c$-axis. } It should be required to involve the interplane
couplings into consideration.

\section{ CONCLUSION}

We have studied the development of the Mn electron spin
correlations in two half-doped Bi-based manganites
Bi$_{0.5}$Sr$_{0.5}$MnO$_{3}$ and Bi$_{0.5}$Ca$_{0.5}$MnO$_{3}$ in
going from the charge-disorder paramagnetic phase to the
charge-ordered antiferromagnetic phase and compared it to
Pr$_{0.5}$Ca$_{0.5}$MnO$_{3}$.

In the Bi-based manganites, the ZF NMR results show that the
unusually large local magnetic field $^{209}H_{loc}$ strongly
favors the dominant $6s^{2}$ character of the lone electron pair
located at the Bi$^{3+}$-ions in both compounds. The mechanism
which connects the $s$ character of the lone pairs of Bi$^{3+}$ to
the high value of $T_{CO}$ is still not clarified.

Compared to Bi$_{0.5}$Sr$_{0.5}$MnO$_{3}$ the small decrease of
$^{209}H_{loc}$ in Bi$_{0.5}$Ca$_{0.5}$MnO$_{3}$ is attributed to
a slight difference in the magnetic structure of the CE$'$-type
spin order presumed for both manganites  rather than to a change
of the wave character of the lone electron pair as it was
suggested for Ca-doped oxide in Ref.~\onlinecite{Garcia_PRB63}.
Our $^{17}$O and $^{209}$Bi ZF NMR data show that, more probably,
the difference in the observed $^{209}H_{loc}$ is due to a
decrease in the canting of the staggered magnetic moments of
Mn$^{3+}$-ions in Bi$_{0.5}$Ca$_{0.5}$MnO$_{3}$ compared to
Bi$_{0.5}$Sr$_{0.5}$MnO$_{3}$. For Bi$_{0.5}$Ca$_{0.5}$MnO$_{3}$
it would be interesting to compare the distribution of staggered
magnetization deduced from NMR data with neutron diffraction data
that are still not available. It would help us to make more
definite conclusion on the role of the stereoactivity of the
Bi$^{3+}$ lone electron pair in the high value of  $T_{CO}$ of the
Bi$_{1-x}$Sr$_{x}$MnO$_{3}$ ($x \leq 0.5$) manganites.

Comparing the evolution of $^{17}$O NMR spectra from CD to CO PM
state, we have demonstrated the qualitative modification of the
spectra at the temperature of order-disorder transition, below
which the localization of itinerant $e_{g}$-electrons manifests
itself in two specific phenomena, namely, the appearance of
interlayer AF spin correlations and the cooperative Jahn-Teller
distortions in the MnO$_{6}$-octahedra sublattice. Indeed, this is
well demonstrated by the proximity of $T_{CO,nmr}$, below which AF
interlayer spin correlations occur, to $T_{crit}$, at which
structural changes occur. This $^{17}$O NMR study demonstrates
that the line corresponding to the apical oxygens is a unique
local tool to study the development of the interlayer Mn-Mn spin
correlations. Indeed for all three manganites this line occurs as
soon as the charge-ordered phase develops on decreasing the
temperature below  $T_{CO}$. It also could be a useful local probe
to study the influence of a magnetic field or an external pressure
on the phase diagram.

In the CO state, $^{17}T_{2}$ is much shorter in the high
frequency part of the spectrum, which corresponds to oxygens in
the $ab$-plane in both Bi-manganites, compared to
Pr$_{0.5}$Ca$_{0.5}$MnO$_{3}$. This great increase of the rate of
the $^{17}$O spin-echo decay can be interpreted in the following
way. The time-dependent fluctuations of the local field can
greatly increase the spin-spin relaxation rates of these oxygens.
The $T_{2}$ effect becomes important when the characteristic
correlation time of the Mn electron spin fluctuations $t_{c}$ is
comparable to $1/\omega$, where $\omega$ is the Larmor precession
of the $^{17}$O nuclear spin in the magnetic field. The maximum of
$^{17}T_{2}^{-1}$ is expected near $t_{c}(T) = 1/\omega$. Thus,
compared with the spin fluctuation spectrum for the CE-ordered
Pr$_{0.5}$Ca$_{0.5}$MnO$_{3}$, in Bi-compounds the spectral
intensity of the Mn electron spin fluctuations is shifted to low
frequencies, comparable to $\omega$.

It is shown on an example of Pr$_{0.5}$Ca$_{0.5}$MnO$_{3}$ and Bi$_{0.5}$Sr$_{0.5}$MnO$_{3}$, that the analysis of the $^{17}$O spectrum imposes some constraints for the theoretical description of the
charge ordered phase in the AF state of manganites. In this analysis
the relative intensity of the low and high frequency $^{17}$O lines is
calculated and compared to the site-centered and bond-centered
models. As the result the fine structure of the $^{17}$O spectrum in
Pr$_{0.5}$Ca$_{0.5}$MnO$_{3}$ is well consistent with predictions of  the site-centered model of CO among the ions (Mn$^{3.5+\delta}$/Mn$^{3.5-\delta}$).
While $^{17}$O NMR analysis of the Mn-Mn spin correlations developed in the AF state of Bi$_{0.5}$Sr$_{0.5}$MnO$_{3}$ indicates on the bond-centered
models of CO as the more appropriate approach for the Bi-based manganite.
\begin{acknowledgments}

We are very grateful to Dr. A.Inyushkin for $^{17}$O isotope
enrichment. The work is supported partly by Russian Foundation for
Basic Research (Grants 02-02-16357a) as well as by CRDF PR 2355.

\end{acknowledgments}

\bibliographystyle{apsrev}
\bibliography{ref}

\end{document}